# Chemical vapor deposition growth of continuous monolayer antiferromagnetic CrOCl films

Chao Chen[1,†], Yulu Liu[1,†], Hongyan Lu[2,†], Zihao Wang[3,†], Bowen Zheng[1], Qian Guo[4,5], Jingkuan Xiao[1], Ping Wang[1], Wanting Xu[1], Yulin Han[2], Mingxuan Chen[1], Xiaofan Cai[1], Jiabei Huang[1], Yaqing Han[1], Di Zhang[1], Renjun Du[1], Alexander S. Mayorov[1,*], Ziying Li[1], Shuai Zhang[1], Yi Huang[1], Tingting Cheng[1], Zhaolong Chen[6], Ronghua Liu[1,7,8], Nujiang Tang[1,7,8], Haibo Ni[9], Di Wu[1], Libo Gao[1,9,10], Xiaoxiang Xi[1], Qianghua Wang[1], Lei Wang[1,9,10,*], Kostya S. Novoselov[3,*] & Geliang Yu[1,9,10,*]

[1]National Laboratory of Solid State Microstructures, Collaborative Innovation Center of Advanced Microstructures, School of Physics, Nanjing University, Nanjing, China.

[2]School of Physics and Physical Engineering, Qufu Normal University, Qufu, China.

[3]Institute for Functional Intelligent Materials, National University of Singapore, Singapore, Singapore.

[4]Department of Physics and Astronomy, University of Manchester, Manchester, UK.

[5]National Graphene Institute, University of Manchester, Manchester, UK.

[6]School of Advanced Materials, Peking University Shenzhen Graduate School, Shenzhen, China.

[7]Jiangsu Provincial Key Laboratory for Nanotechnology, Nanjing University, Nanjing, China.

[8]National Key Laboratory of Spintronics, Nanjing University, Suzhou, China.

[9]International Joint Laboratory on Two Dimensional Materials, Nanjing University, Nanjing, China.

[10]Jiangsu Physical Science Research Center, Nanjing University, Nanjing, China.

[†]These authors contributed equally: Chao Chen, Yulu Liu, Hongyan Lu, Zihao Wang.

[*]E-mail: mayorov@nju.edu.cn; leiwang@nju.edu.cn; kostya@nus.edu.sg; yugeliang@nju.edu.cn

## Abstract

The discovery of two-dimensional magnetic materials has provided an ideal platform for exploring physical phenomena in the two-dimensional limit. However, intrinsic two-dimensional antiferromagnetic materials have been rarely reported, limiting systematic studies of their electronic properties. The discovery of novel intrinsic two-dimensional antiferromagnets and the development of robust synthesis strategies, therefore, remain significant challenges. Here, we report the chemical vapor deposition synthesis of CrOCl monolayer films and nanosheets that exhibit excellent air stability. The CrOCl morphology is tunable, ranging from two-dimensional nanosheets to three-dimensional flower-like structures, with lateral sizes ranging from several microns to continuous monolayer films. Structural characterization confirms the material's composition and high crystalline quality. Furthermore, magnetic measurements, supported by theoretical calculations, reveal a Néel temperature for CrOCl of ≈ 14 K. This work provides a reliable route for preparing two-dimensional antiferromagnetic materials.

# Introduction

Two-dimensional (2D) magnetic materials have recently become a focus of intense research, driven by their intriguing magnetic properties and potential applications in next-generation electronics and optoelectronics[1-5]. For example, $CrI_3$ and $Cr_2Ge_2Te_6$ nanosheets exhibit layer-correlated ferromagnetism at low temperatures[6,7]. Chemical vapor deposition (CVD)-grown $Cr_5Te_8$ exhibits a high Curie temperature, excellent ferromagnetic properties, and a colossal anomalous Hall effect[8-11]. Furthermore, $CrCl_3$ is an A-type antiferromagnet with in-plane magnetic anisotropy[12-14]. Similarly, the magnetic order and anisotropy of related chromium trihalides, such as $CrBr_3$ and $CrI_3$ can be tuned by strain and doping[15-18], making them a versatile platform for studying magnetic quantum phase transitions and designing novel spintronic devices. Additionally, interactions between graphene and antiferromagnetic materials can open a band gap at the Dirac point of graphene and induce phenomena such as the quantum anomalous Hall effect[19,20] and the magnetic proximity effect[21,22].

Although researchers have made great progress in growing thin films of 2D van der Waals materials and 2D magnets, achieving the monolayer limit remains challenging. For instance, rapid, scalable synthesis of epitaxially aligned ferromagnetic ternary metal chalcogenide ($Cr_2Ge_2Te_6$, $Cr_2Si_2Te_6$, $Mn_3Si_2Te_6$) thin films has been reported[23]. Monolayer and few-layer $CrTe_2$ ultrathin films have been synthesized on bilayer graphene by molecular beam epitaxy[24]. However, films obtained by this method are typically several layers thick, greatly hindering the exploration of intrinsic 2D magnetic properties and device applications. CrOCl, an important member of the MOX (M = Sc, Ti, V, Cr, Mn, Fe, Ni, Bi; X = F, Cl, Br) family, is a recently discovered, structurally anisotropic stable 2D ternary compound[25-28], with a Néel temperature of 13.6 K[29], making it a promising antiferromagnet. Previous studies of CrOCl and its heterostructures have relied mainly on mechanical exfoliation[30-34] and chemical vapor transport[29,35]. These methods are complex, time-consuming, unsuited to large-scale production, and cannot guarantee repeatable and consistent sample preparation. In addition, mechanical stacking can degrade crystal quality or leave chemical residue contamination, resulting in weak interfacial coupling[3]. Although CVD-grown CrOCl nanosheets have been demonstrated recently, their thickness and lateral dimensions remain limited[28]. The controllable synthesis of 2D CrOCl monolayer films is therefore of fundamental and technological importance, but to date it has been explored only theoretically[36] and has not yet been realized at a large scale. CVD offers simplicity, high throughput, wafer-scale uniformity, high crystal quality, and is readily scalable[37-42], making it an attractive route to high-quality 2D CrOCl.

In this work, we synthesize high-quality continuous antiferromagnetic CrOCl monolayer films and nanosheets using a CVD method featuring rapid heating and cooling, achieved by moving the furnace, which is crucial for the growth of the CrOCl monolayer films. By adjusting the growth temperature (500-800 °C) and time (1-15 min), the thickness and size of CrOCl samples can be precisely controlled, and the morphology of CrOCl shows a clear transition from 2D nanosheets to three-dimensional (3D) flower-like structures. Notably, a continuous CrOCl monolayer film is obtained at 600 °C with a growth of 5 min. Annular dark-field aberration-corrected scanning transmission electron microscopy (ADF-STEM) images confirm the high crystalline quality of the as-grown CrOCl samples, demonstrating a structure consistent the theoretical simulations. Angle-resolved polarized Raman spectroscopy (ARPRS) and temperature-dependent Raman scattering reveal the structural anisotropy and phase transition behavior of the 2D CrOCl nanosheets. Magnetic measurements, performed using a superconducting quantum interference device (SQUID), show a clear transition in the magnetic susceptibility of CrOCl at 13.9 K. The hysteresis loop of CrOCl gradually diminishes from 3 K to 14 K and eventually disappears, indicating the Néel temperature of CrOCl is approximately 14 K. These experimental finding are supported by density functional theory (DFT) calculations, which predict an antiferromagnetic ground state with a Néel temperature of 13.6 K. Finally, aging experiments on CrOCl nanosheets in an ambient atmosphere demonstrate their excellent environmental stability.

## Results

**Structure and synthesis strategy for 2D CrOCl**

To achieve controlled preparation of continuous CrOCl monolayer films and 2D CrOCl nanosheets with high crystallinity, we used the CVD growth method (Fig. 1a). In this approach, naturally oxidized $CrCl_3$ powder served as the sole source of Cr and O throughout the growth process, thereby simplifying process control and reducing impurities relative to protocols that introduce additional O sources. Importantly, our method exhibits excellent reproducibility and yield in growing CrOCl monolayer films and nanosheets (see Methods for details). CrOCl adopts an orthorhombic layered structure in the $D_{2h}$ (Pmmn) space group, with the Cr-O bilayer sandwiched between two Cl layers stacked along the *c*-axis (Supplementary Fig. 1). Fig. 1b shows a typical optical-microscopy (OM) image of the as-grown CrOCl monolayer film transferred onto a 300-nm-thick $SiO_2$/Si substrate. The flat, continuous film uniformly covers the substrate surface, extending over several hundred microns. Notably, growth time critically determines the film thickness and lateral dimensions of CrOCl, particularly during formation of monolayer films (Supplementary Fig. 2).

When the growth time was increased from 1 to 5 min, isolated CrOCl monolayer domains gradually expanded and coalesced, forming a continuous monolayer film that uniformly covered the substrate (Supplementary Fig. 2a-c). Subsequently, thicker CrOCl nanosheets nucleated on the underlying monolayer and evolved into 3D flower-like CrOCl structures as the growth time was extended from 8 to 15 min (Supplementary Fig. 2d-f). To demonstrate the reproducibility, ten growth runs (Supplementary Fig. 3) carried out under optimal conditions (600 ºC, 5 min) yielded continuous monolayer films with >90% repeatability. Fig. 1c shows a representative optical microscopic (OM) image of CrOCl nanosheets grown on mica, which display a rhombic morphology and a monolayer thickness of 1.2 nm (inset in Fig. 1c). The as-grown nanosheets can also be readily transferred onto a 300-nm-thick $SiO_2$/Si substrate (Fig. 1d), facilitating subsequent characterization. CrOCl grown on *c*-cut sapphire or $SiO_2$/Si likewise exhibits a 3D flower-like morphology (Supplementary Fig. 4). We attribute this morphology to the number of dangling bonds on these substrates (Supplementary Fig. 5d-f); in contrast, the atomically flat surface of mica (Supplementary Fig. 5a-c) is more favorable for the epitaxial growth of 2D CrOCl nanosheets. The atomic force microscopy (AFM) images of CrOCl samples transferred to SiO$_2$/Si (Supplementary Fig. 6) demonstrate a thickness of 1.5 nm, slightly larger than 1.2 nm measured on mica, further supporting the presence of dangling bonds on SiO$_2$. Growth temperature also strongly influences the morphology, thickness, and lateral dimensions of CrOCl. With increasing temperature, the material evolves from 2D shuttle-shaped domain to 2D rhombic flakes and finally to 3D flower-like shapes, accompanied by pronounced lateral growth (Supplementary Fig. 7). AFM measurements (Supplementary Fig. 8) confirm the corresponding thicknesses of nanosheets grown at 550-650 ºC. Because the optical signal of a monolayer is strongly substrate-dependent on mica and SiO$_2$/Si substrates, subsequent analyses focus on few- to thick-layer CrOCl to facilitate interpretation.

Next, the Raman spectra for CVD-grown CrOCl nanosheets of four different thicknesses are presented in Fig. 1e. Three major peaks appear at 207, 414, and 457 cm$^{-1}$, corresponding to $A_g^1$, $A_g^2$, and $A_g^3$ vibrational modes. As the nanosheet thickness increases, the intensity of each peak steadily increases. Calculated $A_g$-mode frequencies differences for monolayer and bulk CrOCl match experiment within 4% (Supplementary Fig. 9). X-ray diffraction (XRD) (Fig. 1f) shows prominent reflections at 11.3°, 34.7°, 46.9°, and 59.6°, indexed to the (001), (003), (200), and (021) planes in agreement with the standard pattern (PDF#28-0371)[29]. The pronounced (001) reflection indicates preferred growth along the [001] direction. X-ray photoelectron spectroscopy (XPS) (Fig. 1g-j) further confirms the bonding environment and composition of the nanosheets. The XPS survey spectrum (Fig. 1g) reveals Cr, O, and Cl species originating from the nanosheets. High-resolution spectra show *Cr 2p$_{3/2}$* and *Cr 2p$_{1/2}$* peaks (Fig. 1h) at 577.7 and 587.3 eV and an *O 1s* peak at 531.5 eV, respectively (Fig. 1i). *Cl 2p$_{3/2}$* and *Cl 2p$_{1/2}$* peaks appear at 201.4 and 199.8 eV, respectively (Fig. 1j). Collectively, these data demonstrate that the CVD-grown CrOCl nanosheets possess high crystalline quality, stoichiometric composition, and scalable synthesis, providing a robust platform for subsequent studies.

**Compositional and structural characterization of CrOCl nanosheets**

To further evaluate the crystal microstructure of CVD-grown CrOCl nanosheets, energy-dispersive X-ray spectroscopy (EDS) and annular dark-field aberration-corrected scanning transmission electron microscopy (ADF-STEM) analyses were performed. From the EDS mapping in Fig. 2a-e, the Cr, O, and Cl signals are uniformly distributed across the sample. The EDS spectrum also confirms the atomic ratio of Cr:O:Cl≈1:1:1 (Fig. 2f). Fig. 2g displays a typical atomic resolution ADF-STEM image of a CrOCl nanosheet, revealing a uniform atomic arrangement with no local discontinuities or defects, thereby demonstrating its high crystal quality. In ADF-STEM mode, the intensity of a nuclear column is approximately proportional to $Z^2$ (where $Z$ is the atomic number), resulting in brighter contrast for Cr columns than for Cl/O columns[43]. We also observe lattice spacings of 0.158 nm for the (020) plane and 0.196 nm for the (200) plane, indicating a highly ordered periodic atomic arrangement. Bulk CrOCl crystallizes in the orthorhombic space group Pmmn (No. 59), in which each Cr column alternates vertically with overlapping Cl/O columns, as illustrated in the structural model (Fig. 2h). A comparable result is obtained for the CrOCl monolayer film (Supplementary Fig. 10), further verifying its high crystal quality. To more precisely elucidate the composition and phase of as-grown nanosheets, we combined quantitative STEM with selected area electron diffraction (SAED) and corresponding structural simulation. Comparing the experimental STEM image (Fig. 2i) with the simulated STEM image (Fig. 2j) shows that the intensities and positions of the Cr, O, and Cl columns coincide. Similarly, SAED patterns from experiment (Fig. 2k) and simulation (Fig. 2l) display an identical single set of quadrilateral diffraction spots, confirming that the nanosheets are single crystals with an orthorhombic structure. Together with the XRD and XPS results, these findings demonstrate that the CVD-grown CrOCl nanosheets possess well-defined composition and crystal structure.

**Orientation distribution and anisotropy of 2D CrOCl**

It is well established that the lattice orientation of 2D materials is strongly dependent on the growth substrate[44,45]. According to previous work, the mica substrate possesses a six-fold symmetrical structure (Fig. 3a)[46], which guides the oriented growth of 2D materials. Fig. 3b displays an OM image of CVD-grown CrOCl nanosheets on mica. As expected, the CrOCl nanosheets are preferentially oriented at three principal angles (0°, 120°, -120°), as confirmed by the statistical results in Fig. 3c. Another notable factor is that the vdW interaction between CrOCl and mica substrate plays a critical role in the growth process. The crystal orientation and anisotropy of 2D CrOCl were investigated using angle-resolved polarization Raman spectroscopy (ARPRS). Fig. 3d shows an OM image of a CrOCl nanosheet used for ARPRS measurements. The results demonstrate pronounced intensity differences among the three Raman modes ($A_g^1$, $A_g^2$, and $A_g^3$) of CrOCl nanosheets under parallel (XX) and perpendicular (XY) configurations (Fig. 3e). All three modes show periodic variations from 0° to 180° (Supplementary Fig. 11). In the XX configuration, the $A_g^1$, $A_g^2$, and $A_g^3$ modes exhibit two-fold symmetry with a 180° period (Fig. 3f-h). In this configuration, $A_g^1$ and $A_g^3$ reach maxima at 60° and 240°, while $A_g^2$ mode shows maxima at 150° and 330°. This behavior may reflect phonon-induces modulation of electronic states. The *x*-polarized photon-electron interaction appears to be dominated by $A_g^2$, whereas $A_g^1$ and $A_g^3$ couple more strongly under *y*-polarization[29]. In the XY configuration, all $A_g$ modes display four-fold symmetry with a 90° period, peaking at 15°, 105°, 195°, and 285° (Fig. 3i-k). These findings reveal a close correlation between the crystal orientation of CrOCl and its Raman response.

**Temperature-dependent Raman properties of 2D CrOCl nanosheets**

To further explore the phase transition behavior of 2D CrOCl nanosheets, we performed temperature-dependent Raman scattering measurements on samples supported by a 300-nm-thick $SiO_2$/Si substrate. Fig. 4a shows three distinct peaks at ~207, 414, and 457 cm$^{-1}$, corresponding to $A_g^1$, $A_g^2$, and $A_g^3$ modes, respectively, that persist throughout 8.3-300 K. When the temperature is lowered from 300 K to 8.3 K, all three peaks exhibit a blue-shift, more clearly visible in their magnified spectra (Supplementary Fig. 12a-c). To clarify the evolution of peak intensity, we plotted Raman spectra from 8.3 K to 300 K using a common baseline (Fig. 4b) and extracted the temperature dependence of the $A_g^1$, $A_g^2$, and $A_g^3$ intensities (Fig. 4c-e). Above 27 K, the intensity first increases and then decreases, whereas below 27 K it remains nearly constant. The temperature dependence of the corresponding frequency shifts likewise and shows a pronounced inflection near 27 K (Supplementary Fig. 12d-f). Taken together, these observations indicate that CrOCl undergoes a phase transition at ~27 K.

**Magnetic properties of CrOCl nanosheets**

Next, we investigated the magnetic properties of CVD-grown CrOCl nanosheets using a superconducting quantum interference device vibrating-sample magnetometer (SQUID-VSM). Because SQUID measurements require a large mass, we stacked multiple nanosheets on a single $SiO_2$/Si substrate to obtain a sufficiently strong antiferromagnetic signal. Fig. 5a shows a schematic of the SQUID measurement, with the magnetic field aligned parallel to the *c*-axis. Out-of-plane hysteresis loops recorded at various temperatures (Fig. 5b) reveal switchings between ±3~6 T at 3 K, characteristic of antiferromagnetism. With increasing temperature, the hysteresis loop progressively tightens and disappears at 20 K. The temperature dependence of Δmoment = ($moment_{max}$ - $moment_{min}$) and Δmagnetic field = (magnetic $field_{max}$ - magnetic $field_{min}$) (Supplementary Fig. 13) likewise decreases monotonically, vanishing at 20 K. Figure 5c plots the magnetic susceptibility of CrOCl versus temperature at 0.1 T field applied parallel and perpendicular to the *c*-axis, yielding a Néel temperature of 13.9 K, in excellent agreement with the theoretical calculation (Fig. 5d). Fig. 5e presents a top view of CrOCl crystal structure and Fig. 5f depicts the antiferromagnetic spin arrangement calculated on an unshifted **k** mesh of 13×15×7 at 0 K. Finally, the Raman spectra collected immediately after growth and after eight months of air exposure are nearly identical (Supplementary Fig. 14), confirming the outstanding air stability of CrOCl and underscoring its promise for studies and applications under ambient conditions.

**Discussion**

In summary, we synthesized high-quality continuous CrOCl monolayer films and nanosheets on mica by the CVD method. Precise control of the growth temperature and duration systematically tunes the morphology, thickness, and lateral size of the CrOCl samples. Increasing the growth temperature or extending the growth time drives a transition from 2D nanosheets to 3D flower-like structures within a narrow parameter window, underscoring the scalability of the process and enabling morphology-specific applications. ADF-STEM, corroborated by structural simulations, confirms the high crystallinity of both monolayer films and nanosheets. ARPRS reveals pronounced polarization dependence and strong in-plane anisotropy. Temperature-dependent Raman scattering identifies a structural transition at ~27 K. SQUID magnetometry demonstrates that the hysteresis loop narrows from 3 K to 14 K and disappears by ~20 K, whereas magnetic susceptibility data yield a Néel temperature of 13.9 K. DFT calculations likewise predict an antiferromagnetic ground state, with a Néel temperature of ~13.6 K. The nanosheets also exhibit excellent air stability. Collectively, these findings position CrOCl as a versatile platform for two-dimensional magnetism and spintronic devices.

## Methods

**Growth and transfer of CrOCl monolayer film and CrOCl nanosheets**

High-quality, continuous CrOCl monolayer films and CrOCl nanosheets were synthesized on mica substrates via a conventional CVD method. Naturally oxidized $CrCl_3$ powders (5 mg, Alfa, 99.9%) as the sole precursor were placed in a quartz boat positioned at the center of the hot zone, maintained at temperatures between 500-800 °C. A freshly peeled 1 cm×1 cm mica ($KMg_3AlSi_3O_{10}F_2$) substrate was placed face-down over the $CrCl_3$ powder. Before growth, the reaction chamber was purged with 500 sccm of argon (Ar, 99.999%) for 10 min. Subsequently, the furnace was heated to the growth temperature under a 100 sccm Ar flow and maintained for 1-15 min to grow the CrOCl monolayer films and CrOCl nanosheets. Here, the growth time, growth temperature, and the temperature gradient during the cooling process are crucial for growing of the CrOCl samples. Through regulating the growth parameters, optimal growth conditions and a CrOCl monolayer film with a maximum size of approximately 0.5 cm² can be obtained (see Supplementary Table 1-4). For further characterization, the as-synthesized CrOCl samples were transferred onto $SiO_2$/Si (300 nm) substrates or copper grids by a polymethyl methacrylate (PMMA)-assisted transfer method.

**Structure and composition characterization**

The morphology, phase structure, and thickness of CVD-grown CrOCl samples were characterized by optical microscopy (OM) (ECLIPSE LV150N, Nikon), scanning electron microscopy (SEM) (Carl Zeiss, GeminiSEM 500), X-ray diffraction (XRD) (Rigaku D/MAX-Ultima III, Japan), and atomic force microscopy (AFM) (Bruker Dimension Fastscan system).

Room-temperature Raman spectra were obtained on a WITec/alpha 300R confocal microscope with a 532 nm laser excitation source. Angle-resolved polarization Raman spectroscopy (ARPRS) and temperature-dependent Raman spectra were collected using an Acton SpectraPro SP-2300 coupled with a CRYOSTATION C2 system, employing an 1800 grooves/mm grating. For ARPRS measurements, the incident polarization was varied by rotating a half-wave plate. Two polarization configurations were recorded by aligning the scattering light polarization parallel or perpendicular to the incident polarization direction. The X-ray photoelectron spectroscopy (XPS) analysis was performed on a PHI 5000 VersaProbe spectrometer using a monochromatic Al K(alpha) X-ray source. The atomic structure and element distribution of CrOCl were evaluated via high-angle annular dark-field scanning transmission electron microscopy (ADF-STEM) imaging and energy-dispersive X-ray spectroscopy (EDS) mapping using a probe aberration-corrected STEM (Titan cubed G2 60-300) at 300 kV.

**Magnetic property measurements**

SQUID-VSM measurements were conducted using a Quantum Design magnetic property measurement system. CrOCl samples on $SiO_2$/Si (300 nm) substrates were tested at temperatures ranging from 3 to 20 K under an out-of-plane magnetic field varying from -6.3 to 6.3 T. The temperature-dependent magnetic susceptibility of CrOCl samples on $SiO_2$/Si (300 nm) substrates were tested at temperatures ranging from 3 to 300 K under a 0.1 T out-of-plane and in-plane magnetic field.

**Computational details**

To theoretically investigate the magnetic properties and Raman spectra of CrOCl, first-principles calculations within the framework of density functional theory (DFT)[47] are performed with the Vienna ab initio Simulation Package (VASP)[48,49]. The Perdew-Burke-Ernzerhof (PBE) parametrized generalized gradient approximation (GGA) is employed to describe the exchange-correlation potentials[50,51]. The projector-augmented-wave (PAW) pseudopotentials[52] are utilized to model the electron-ion interactions. The kinetic energy cutoff and the charge density cutoff are set to 500 eV. The charge densities are determined self-consistently on an unshifted **k**-mesh of 13 × 15 × 7 and 13 × 15 × 1 for geometric structures calculations of bulk and monolayer CrOCl, respectively. The convergence criteria were $1\times10^{-6}$ eV for the energy difference in electronic self-consistent calculation and $1\times10^{-2}$ eV/Å for the residual forces on atoms. To better describe the strong correlation effect of $d$ electrons in Cr, the DFT + $U$ method is used for all calculations, with the Hubbard's $U$ and Hund's $J$ for the Cr $d$ electron setting to 3.0 and 1.0 eV, respectively, which were used in previous studies[27], and are comparable to those adopted in modelling CrSCl[53] and CrI$_3$[54]. The density functional dispersion correction (D3-Grimme) is adopted for the van der Waals (vdW) interactions[55]. The Néel temperature is determined from Monte Carlo simulations based on the heat bath algorithms[56,57].

## Data availability

The data used in this study are available from the corresponding authors upon request.

# References

1. Huang, B. et al. Emergent phenomena and proximity effects in two-dimensional magnets and heterostructures. *Nat. Mater.* **19**, 1276 (2020).
2. Hossain, M., Qin, B., Li, B. & Duan, X. Synthesis, characterization, properties and applications of two-dimensional magnetic materials. *Nano Today* **42**, 101338 (2022).
3. Yao, R. et al. Controlled synthesis of 2D ferromagnetic/antiferromagnetic $Cr_7Te_8$/MnTe vertical heterostructures for high-tunable coercivity. *ACS Nano* **18**, 23508 (2024).
4. Blundo, E. et al. Localisation-to-delocalisation transition of moiré excitons in $WSe_2$/$MoSe_2$ heterostructures. *Nat. Commun.* **15**, 1057 (2024).
5. Huang, S. et al. Giant magnetoresistance induced by spin-dependent orbital coupling in $Fe_3GeTe_2$/graphene heterostructures. *Nat. Commun.* **16**, 2866 (2025).
6. Huang, B. et al. Layer-dependent ferromagnetism in a van der Waals crystal down to the monolayer limit. *Nature* **546**, 270 (2017).
7. Gong, C. et al. Discovery of intrinsic ferromagnetism in two-dimensional van der Waals crystals. *Nature* **546**, 265 (2017).
8. Chen, C. et al. Air-stable 2D $Cr_5Te_8$ nanosheets with thickness-tunable ferromagnetism. *Adv. Mater.* **34**, 2107512 (2022).
9. Tang, B. et al. Phase engineering of $Cr_5Te_8$ with colossal anomalous Hall effect. *Nat. Electron.* **5**, 224 (2022).
10. Fu, L., Llacsahuanga Allcca, A. E. & Chen, Y. P. Coexisting ferromagnetic-antiferromagnetic state and giant anomalous Hall effect in chemical vapor deposition-grown 2D $Cr_5Te_8$. *ACS Nano* **18**, 33381 (2024).
11. Jiang, Q. et al. Controlled growth of submillimeter-scale $Cr_5Te_8$ nanosheets and the domain wall nucleation governed magnetization reversal process. *Nano Lett.* **24**, 1246 (2024).
12. Klein, D. R. et al. Enhancement of interlayer exchange in an ultrathin two-dimensional magnet. *Nat. Phys.* **15**, 1255 (2019).
13. Wang, Z. et al. Determining the phase diagram of atomically thin layered antiferromagnet $CrCl_3$. *Nat. Nanotechnol.* **14**, 1116 (2019).
14. Wang, J. et al. Physical vapor transport growth of antiferromagnetic $CrCl_3$. *Adv. Sci.* **10**, 2203548 (2023).
15. Jiang, S. W., Li, L. Z., Wang, Z. F., Mak, K. F. & Shan, J. Controlling magnetism in 2D $CrI_3$ by electrostatic doping. *Nat. Nanotechnol.* **13**, 549 (2018).
16. Huang, B. et al. Electrical control of 2D magnetism in bilayer $CrI_3$. *Nat. Nanotechnol.* **13**, 544 (2018).
17. Jiang, S., Li, L., Wang, Z., Shan, J. & Mak, K. F. Spin tunnel field-effect transistors based on two-dimensional van der Waals heterostructures. *Nat. Electron.* **2**, 159 (2019).
18. Webster, L. & Yan, J.-A. Strain-tunable magnetic anisotropy in monolayer $CrCl_3$, $CrBr_3$, and $CrI_3$. *Physical Review B* **98**, 144411 (2018).
19. Qiao, Z. et al. Quantum anomalous Hall effect in graphene proximity coupled to an antiferromagnetic insulator. *Physical Review Letters* **112**, 116404 (2014).
20. Högl, P. et al. Quantum anomalous Hall effects in graphene from proximity-induced uniform and staggered spin-orbit and exchange coupling. *Physical Review Letters* **124**, 136403 (2020).
21. Yang, B. et al. Electrostatically controlled spin polarization in graphene-CrSBr magnetic proximity heterostructures. *Nat. Commun.* **15**, 4459 (2024).
22. Ghiasi, T. S. et al. Quantum spin Hall effect in magnetic graphene. *Nat. Commun.* **16**, 5336 (2025).
23. Giri, A. et al. Large-area epitaxial film growth of van der Waals ferromagnetic ternary chalcogenides. *Adv. Mater.* **33**, 2103609 (2021).

24. Zhang, X. et al. Room-temperature intrinsic ferromagnetism in epitaxial $CrTe_2$ ultrathin films. *Nat. Commun.* **12**, 2492 (2021).

25. Miao, N., Xu, B., Zhu, L., Zhou, J. & Sun, Z. 2D intrinsic ferromagnets from van der Waals antiferromagnets. *J. Am. Chem. Soc.* **140**, 2417 (2018).

26. Nair, A. K., Rani, S., Kamalakar, M. V. & Ray, S. J. Bi-stimuli assisted engineering and control of magnetic phase in monolayer CrOCl. *Physical Chemistry Chemical Physics* **22**, 12806 (2020).

27. Gu, P. et al. Magnetic phase transitions and magnetoelastic coupling in a two-dimensional stripy antiferromagnet. *Nano Lett.* **22**, 1233 (2022).

28. Tang, Y. et al. Synthesis of highly anisotropic 2D insulator CrOCl nanosheets for interfacial symmetry breaking in isotropic 2D semiconductors. *Adv. Mater.* **37**, 2405358 (2024).

29. Zhang, T. et al. Magnetism and optical anisotropy in van der Waals antiferromagnetic insulator CrOCl. *ACS Nano* **13**, 11353 (2019).

30. Wei, Z. et al. Van der Waals interlayer coupling induces distinct linear dichroism in $WSe_2$ photodetectors. *Adv. Opt. Mater.* **11**, 2201962 (2023).

31. Zhang, T. et al. Tuning the exchange bias effect in 2D van der Waals ferro-/antiferromagnetic $Fe_3GeTe_2$/CrOCl heterostructures. *Adv. Sci.* **9**, 2105483 (2022).

32. Zheng, X. et al. Symmetry engineering induced in-plane polarization in $MoS_2$ through van der Waals interlayer coupling. *Adv. Funct. Mater.* **32**, 2202658 (2022).

33. Liu, D. et al. Tunable van der Waals doping in $WS_2$/CrOCl heterostructure by interlayer coupling engineering. *ACS Appl. Electron. Mater.* **5**, 3973 (2023).

34. Cao, S. et al. Magnetic-electrical synergetic control of non-volatile states in bilayer graphene-CrOCl heterostructures. *Adv. Mater.* **37**, 2411300 (2024).

35. Zeng, Y. et al. 2D FeOCl: a highly in-plane anisotropic antiferromagnetic semiconductor synthesized via temperature-oscillation chemical vapor transport. *Adv. Mater.* **34**, 2108847 (2022).

36. Jang, S. W. et al. Hund's physics and the magnetic ground state of CrOX (X=Cl,Br). *Physical Review Materials* **5**, 034409 (2021).

37. Zhou, J. et al. Composition and phase engineering of metal chalcogenides and phosphorous chalcogenides. *Nat. Mater.* **22**, 450 (2023).

38. Zhou, J. et al. A library of atomically thin metal chalcogenides. *Nature* **556**, 355 (2018).

39. Li, B. et al. Van der Waals epitaxial growth of air-stable $CrSe_2$ nanosheets with thickness-tunable magnetic order. *Nat. Mater.* **20**, 818 (2021).

40. Zhang, K. et al. Epitaxial substitution of metal iodides for low-temperature growth of two-dimensional metal chalcogenides. *Nat. Nanotechnol.* **18**, 448 (2023).

41. Zhou, J. et al. Heterodimensional superlattice with in-plane anomalous Hall effect. *Nature* **609**, 46 (2022).

42. Bian, M. et al. Dative epitaxy of commensurate monocrystalline covalent van der Waals moiré supercrystal. *Adv. Mater.* **34**, 2200117 (2022).

43. Sun, R., Wang, Z., Saito, M., Shibata, N. & Ikuhara, Y. Atomistic mechanisms of nonstoichiometry-induced twin boundary structural transformation in titanium dioxide. *Nat. Commun.* **6**, 7120 (2015).

44. Shoaib, M. et al. Directional growth of ultralong $CsPbBr_3$ perovskite nanowires for high-performance photodetectors. *J. Am. Chem. Soc.* **139**, 15592 (2017).

45. Wang, Y. et al. Temperature difference triggering controlled growth of all-inorganic perovskite nanowire arrays in air. *Small* **14**, 1803010 (2018).

46. Ma, Z. et al. Chemical vapor deposition growth of high crystallinity $Sb_2Se_3$ nanowire with strong anisotropy for near‐infrared photodetectors. *Small* **15**, 1805307 (2019).
47. Baroni, S., de Gironcoli, S., Dal Corso, A. & Giannozzi, P. Phonons and related crystal properties from density-functional perturbation theory. *Reviews of Modern Physics* **73**, 515 (2001).
48. Kresse, G. & Furthmüller, J. Efficiency of ab-initio total energy calculations for metals and semiconductors using a plane-wave basis set. *Computational Materials Science* **6**, 15 (1996).
49. Kresse, G. & Furthmüller, J. Efficient iterative schemes for ab initio total-energy calculations using a plane-wave basis set. *Physical Review B* **54**, 11169 (1996).
50. Schlipf, M. & Gygi, F. Optimization algorithm for the generation of ONCV pseudopotentials. *Computer Physics Communications* **196**, 36 (2015).
51. Hamann, D. R. Optimized norm-conserving Vanderbilt pseudopotentials. *Physical Review B* **88**, 085117 (2013).
52. Blöchl, P. E. Projector augmented-wave method. *Physical Review B* **50**, 17953 (1994).
53. Wang, C. et al. A family of high-temperature ferromagnetic monolayers with locked spin-dichroism-mobility anisotropy: MnNX and CrCX (X = Cl, Br, I; C = S, Se, Te). *Science Bulletin* **64**, 293 (2019).
54. Jiang, P. et al. Stacking tunable interlayer magnetism in bilayer $CrI_3$. *Physical Review B* **99**, 144401 (2019).
55. Grimme, S., Antony, J., Ehrlich, S. & Krieg, H. A consistent and accurate ab initio parametrization of density functional dispersion correction (DFT-D) for the 94 elements H-Pu. *The Journal of Chemical Physics* **132**, (2010).
56. Metropolis, N. & Ulam, S. The Monte Carlo method. *Journal of the American Statistical Association* **44**, 335 (1949).
57. Miyatake, Y., Yamamoto, M., Kim, J. J., Toyonaga, M. & Nagai, O. On the implementation of the 'heat bath' algorithms for Monte Carlo simulations of classical Heisenberg spin systems. *Journal of Physics C: Solid State Physics* **19**, 2539 (1986).

## Acknowledgements

The authors would like to thank the International Joint Lab of 2D Materials at Nanjing University for the support, and Yuanchen Co, Ltd (http://www.monosciences.com) for the High-Quality 2D Material Transfer System. G.Y. acknowledges the financial support from the National Key R&D Program of China (Nos. 2024YFB3715400, Nos. 2022YFA120470, 2021YFA1400400), the National Natural Science Foundation of China (No. 11974169), the Natural Science Foundation of Jiangsu Province (No. BK20233001), and the support form Nanjing University International Collaboration Initiative. L.W. acknowledges the National Key Projects for Research and Development of China (Nos. 2022YFA120470, 2021YFA1400400), National Natural Science Foundation of China (No. 12074173), Natural Science Foundation of Jiangsu Province (No. BK20220066 and BK20233001). K.S.N. is grateful to the Ministry of Education, Singapore (Research Centre of Excellence award to the Institute for Functional

Intelligent Materials, I-FIM, project No. EDUNC-33-18-279-V12) and to the Royal Society (UK, grant number RSRP\ R\190000) for support. R.D. acknowledges the grant from the National Natural Science Foundation of China (No. 12004173).

**Author contributions**

G.Y. conceived the work, and C.C designed and synthesized the samples. Y.L. completed the Raman spectra measurements under the supervision of X.X. H.L., Yu.H., and Q.W. provided the theoretical calculation. C.C., Z.W., and Q.G. optimized the experimental details. B.Z. performed the SQUID measurements under the supervision of D.W. J.X., W.X., and M.C. helped the data analysis. X.C. and J.H. carried out the AFM measurements. P.W., D.Z., and Ya.H. contributed to the SEM measurements. Z.L., S.Z., and Yi.H. helped the SQUID measurement under the supervision of R.L. and N.T. T.C., Z.C., and H.N. supported the XRD measurements. C.C. performed the data analysis under the suggestion of R.D., A.S.M., L.W., and L.G. C.C. prepared the manuscript with input from L.W., K.S.N., and G.Y.

**Competing interests**

The authors declare no competing interests.

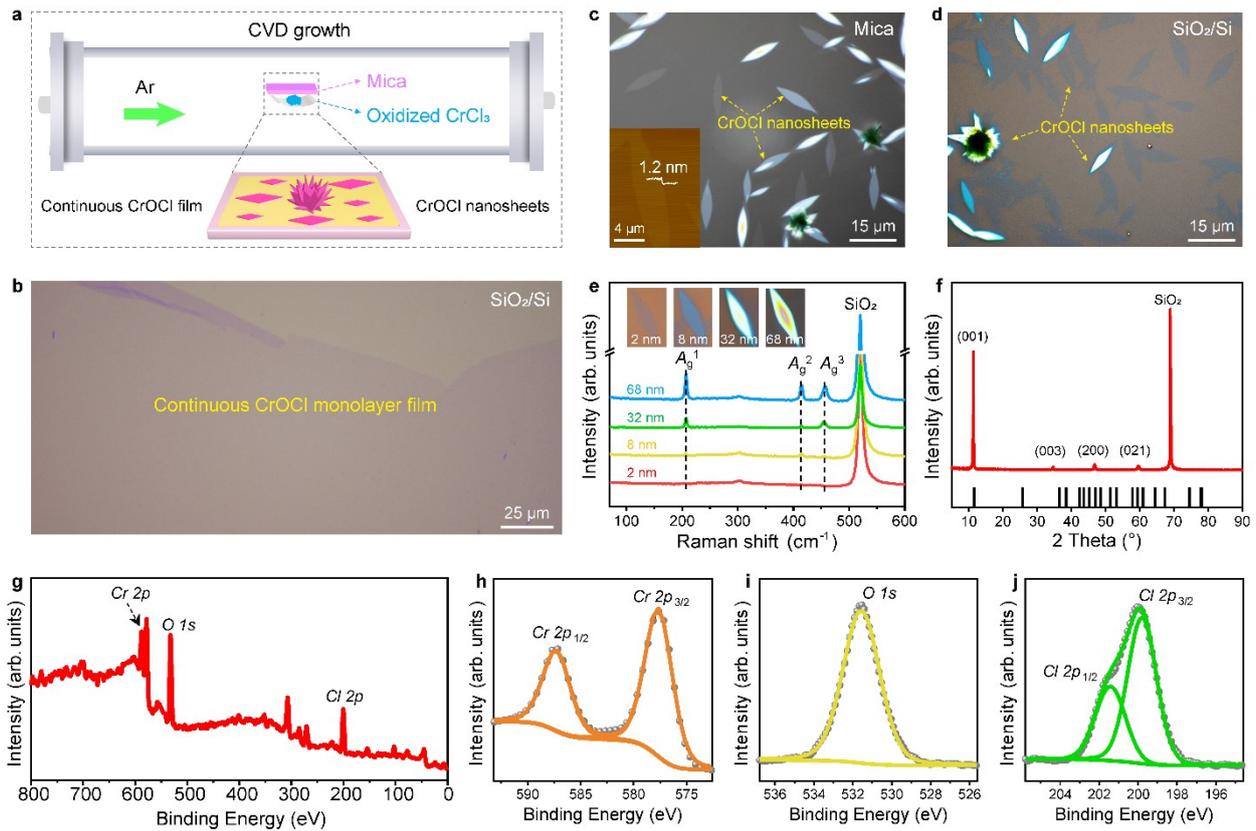

**Fig. 1 | Structural characterizations and synthesis strategy of two-dimensional (2D) CrOCl. a** Schematic diagram of the chemical vapor deposition (CVD) growth of continuous CrOCl monolayer films and CrOCl nanosheets. The green arrow indicates the direction of argon (Ar) gas. **b** A typical optical microscopic (OM) image of as-grown CrOCl monolayer film transferred onto SiO$_2$/Si (300 nm) substrate. OM image of synthesized CrOCl nanosheets (**c**) on a mica substrate and (**d**) transferred to SiO$_2$/Si (300 nm) substrate, inset is the corresponding atomic force microscopy (AFM) image. **e** Raman spectra of as-grown CrOCl nanosheets with different thicknesses on SiO$_2$/Si (300 nm). **f** X-ray diffraction (XRD) of as-grown CrOCl nanosheets on SiO$_2$/Si (300 nm). **g** X-ray photoelectron spectroscopy (XPS) survey spectra and detailed signals of the (**h**) *Cr 2p*, (**i**) *O 1s*, and (**j**) *Cl 2p* of CrOCl nanosheets. The dots and lines represent experimental data and fitted curves, respectively.

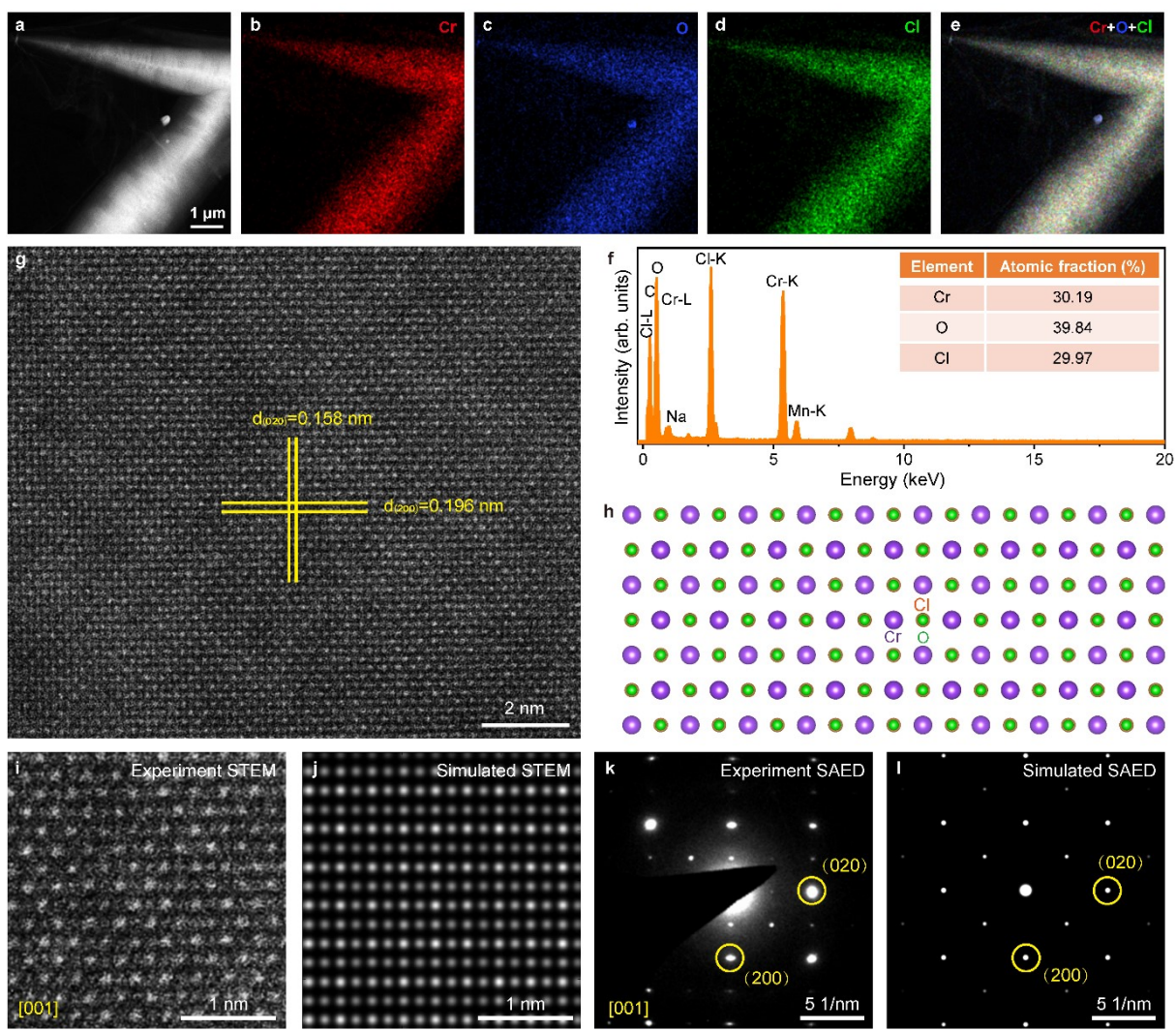

**Fig. 2 | Compositional and structural characterization of CrOCl nanosheets. a** Low-magnification annular dark-field aberration-corrected scanning transmission electron microscopy (ADF-STEM) image of a few-layer CrOCl nanosheets. **b-e** Energy-dispersive X-ray spectroscopy (EDS) mapping of Cr, O, Cl elements and corresponding overlap of CrOCl nanosheets. **f** EDS spectra of chemical vapor deposition (CVD)-grown CrOCl nanosheets. **g** High-resolution ADF-STEM image of the CrOCl grain and (**h**) corresponding atomic model of the CrOCl crystal. The two pairs of horizontal and vertical yellow solid lines represent the (200) plane and (020) plane respectively in (**g**). The purple, green and orange represent the Cr, O and Cl atoms respectively in (**h**). **i,j** Experiment and simulated ADF-STEM images of the CrOCl crystal along [001] direction. **k,l** Experiment and simulated selected area electron diffraction (SAED) patterns of the CrOCl crystal along [001] direction. The yellow circles correspond to the (200) plane and (020) plane respectively.

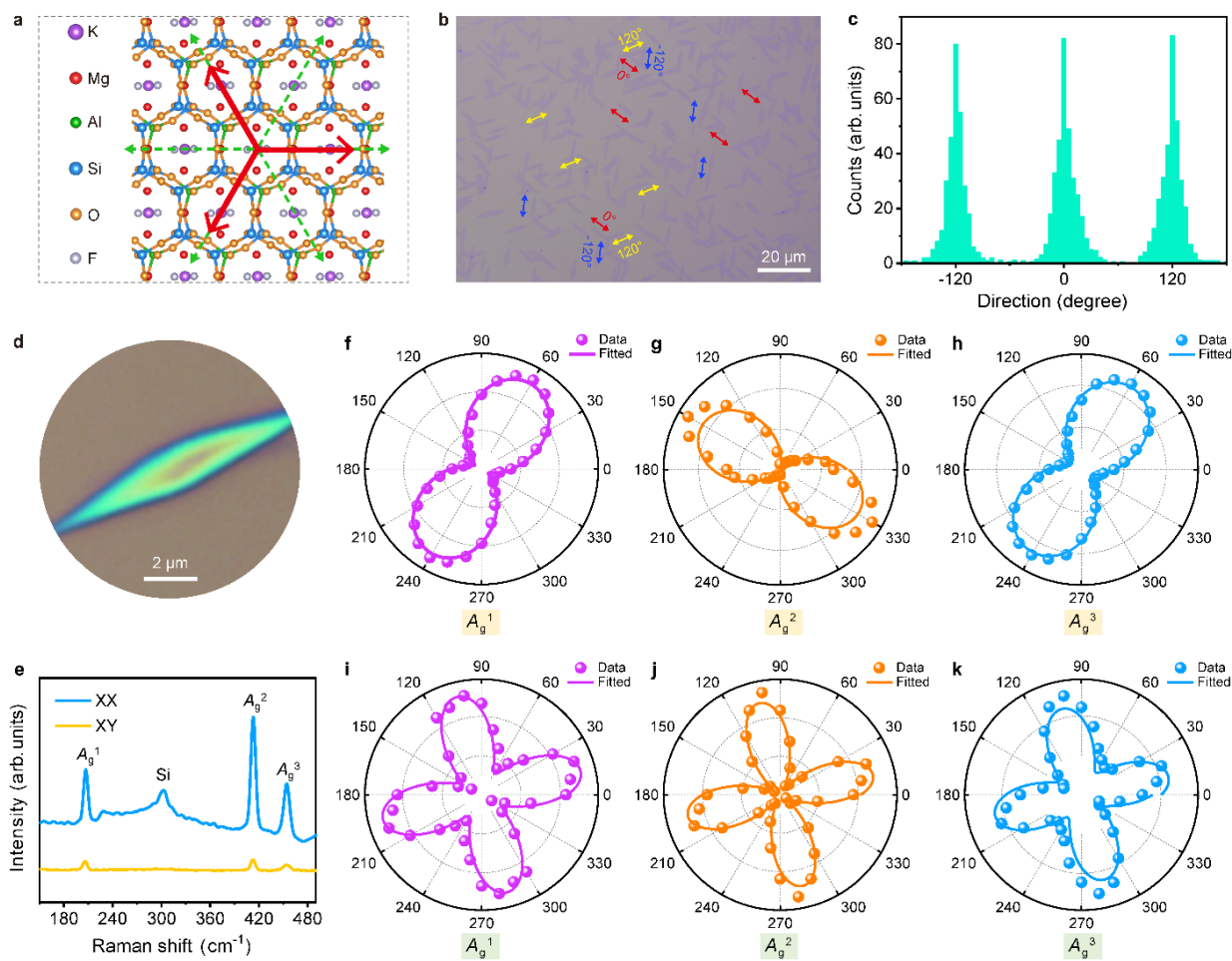

**Fig. 3 | Orientation distribution and angle-resolved polarized Raman spectroscopy (ARPRS) anisotropic characterization of two-dimensional (2D) CrOCl. a** Atomic structure of the mica substrate showing sixfold symmetry (green dotted arrow). Three red solid arrows represent the growth direction of the sample. **b** Optical microscopic (OM) image of CrOCl nanosheets grown on mica, with the three primary orientations marked by yellow, blue, and red arrows. **c** Statistical orientation distribution of CrOCl nanosheets on mica. **d** OM image of a typical CrOCl thin flake on a SiO$_2$/Si (300 nm) substrate. **e** Polarized Raman spectra in parallel (XX) and perpendicular (XY) configurations. **f-k** Polar plots of Raman intensity for $A_g^1$, $A_g^2$, and $A_g^3$ modes under (**f-h**) parallel and (**i-k**) perpendicular configurations. The dots and lines represent experimental data and fitted curves, respectively.

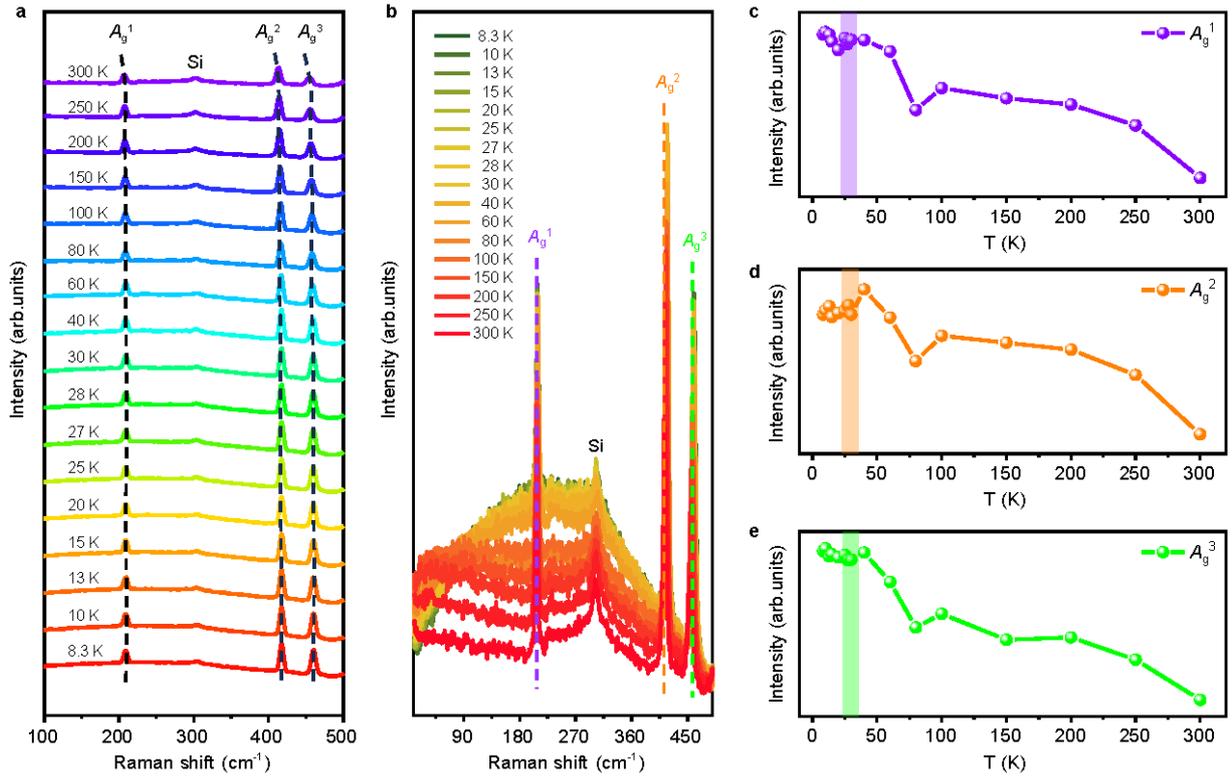

**Fig. 4 | Temperature-dependent Raman spectra of two-dimensional (2D) CrOCl nanosheets. a** Raman spectra of CrOCl at different temperatures (8.3~300 K). **b** Raman spectra of CrOCl at 8.3~300 K with the same baseline. Temperature dependency of Raman intensity for the (**c**) $A_g^1$ (purple), (**d**) $A_g^2$ (orange), and (**e**) $A_g^3$ (green) modes. Three shaded bars represent the transition temperatures in (**c-e**).

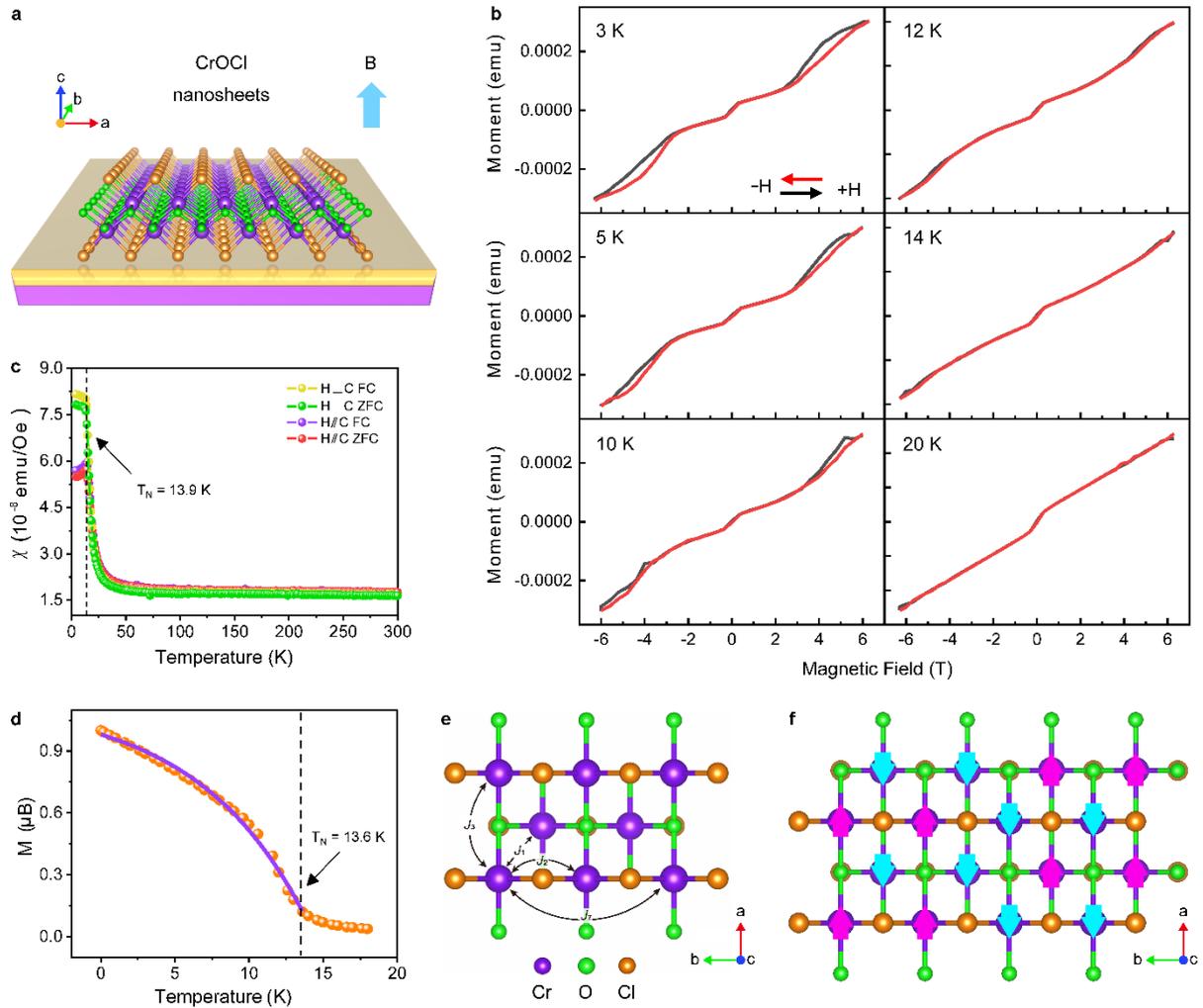

**Fig. 5 | Magnetic property characterization of CrOCl nanosheets. a** Superconducting quantum interference device (SQUID) measurement schematic of CrOCl nanosheets. The magnetic field direction is upward (blue wide arrow). **b** Hysteresis loops measured at different temperatures for CrOCl nanosheets. **c** The magnetic susceptibility ($\chi$) as a function of temperature for CrOCl in a 0.1 T magnetic field parallel and perpendicular to the *c*-axis. **d** Calculated Néel temperature of the bulk CrOCl. **e** The top view of CrOCl crystal structure. $J_n$ refers to the *n*th neighbor coupling constant. **f** The antiferromagnetic (AFM) spin order on an unshifted **k**-mesh of 13 × 15 × 7 under 0 K. The up (magenta) and down (blue) arrows represent the spin direction.